%
%
%

\font\llbf=cmbx10 scaled\magstep2
\font\lbf=cmbx10 scaled\magstep1

\def\ni{\noindent}

\def\edth{{\hskip 3pt {}^{\prime } \kern -6pt \partial }} 

\def\uA{\underline A\,}
\def\uB{\underline B\,}

\def\ua{\underline a\,}

\def\bi{\bf i\,}

\baselineskip 14pt plus 1pt


\ni
{\llbf On a class of 2-surface observables in general relativity}

\bigskip
\ni
{\bf L\'aszl\'o B. Szabados}

\ni
Research Institute for Particle and Nuclear Physics

\ni
H-1525 Budapest 114, P.O.Box 49, Hungary

\ni
E-mail: lbszab@rmki.kfki.hu

\bigskip
\bigskip
\ni
The boundary conditions for canonical vacuum general relativity is 
investigated at the quasi-local level. It is shown that fixing the 
{\it area element on the 2-surface ${\cal S}$} (rather than the 
induced 2-metric) is enough to have a well defined constraint algebra, 
and a well defined Poisson algebra of basic Hamiltonians parameterized 
by shifts that are tangent to {\it and divergence free on ${\cal S}$}. 
The evolution equations preserve these boundary conditions, and the 
value of the basic Hamiltonians gives 2+2--covariant, gauge-invariant 
2-surface observables. The meaning of these observables is also 
discussed.

\bigskip
\bigskip

\ni
{\lbf 1 Introduction}
\medskip
\ni
As is well known, in a spacetime that is asymptotically flat at 
spatial infinity the ten classical conserved quantities, viz. the 
energy-momentum and relativistic angular momentum (i.e. including 
the centre-of-mass), can be introduced in several different ways. 
One possibility is to use a canonical/Hamiltonian approach [1-4]. 
However, to have a deeper understanding e.g. of the (geometrical 
or thermodynamical) properties of black holes, for example their 
entropy, the conserved quantities, or, more generally, the 
observables of the gravitational `field' must be introduced at the 
{\it quasi-local} level. Such investigations lead to the so-called 
surface degrees of freedom [5-8], and to the large variety of 
proposals for the quasi-local energy-momentum and angular momentum [9]. 
A further motivation of searching for quasi-local observables is the 
remarkable result that all the global observables for the vacuum 
gravitational field in a closed universe, built as spatial integrals 
of local functions of the initial data and their derivatives, are 
necessarily vanishing [10,11]. Thus in closed universes we can 
associate non-trivial, locally constructible observables only with 
subsystems, bounded by some closed spacelike 2-surface. 

The aim of the present note is to discuss certain quasi-local, 
2-surface observables within the framework of canonical vacuum general 
relativity. Although in the literature there is a nice and quite 
general analysis using explicit background structures (see e.g. 
[12,13]), here we follow a more traditional (and perhaps more 
`pedestrian') approach, and no such background structure will be used. 
In the subsequent analysis, in addition to the functional 
differentiability of various functions on the phase space (due to Regge 
and Teitelboim [1]), three new requirements, already appeared in the 
asymptotically flat context [2-4,14], will be expected to be satisfied 
at the quasi-local level: 
a. The evolution equations should preserve the boundary conditions 
(i.e. the boundary conditions should be compatible with the evolution 
equations); 
b. The Hamiltonians, and hence, in particular, the constraints, should 
close to a Poisson algebra; 
c. The value of the Hamiltonian on the constraint surface should be 
a 2+2--covariant, gauge invariant observable. 

We show that the observables introduced in [5-8] are well defined even 
under much weaker boundary conditions. It will be shown that 
1. fixing the {\it area element on the 2-surface ${\cal S}$} rather 
than the induced 2-metric is enough to have 
i. a well defined constraint algebra ${\cal C}$, and 
ii. a well defined Poisson algebra ${\cal H}_0$ of basic Hamiltonians 
parameterized by shifts that are tangent to ${\cal S}$ and {\it 
divergence free with respect to the intrinsic Levi-Civita connection 
on ${\cal S}$}. 
2. The evolution equations preserve these boundary conditions; and 
3. the value of the basic Hamiltonians give 2+2--covariant, 
gauge-invariant 2-surface observables. 

In the next section the basic variational formula of the constraints 
is recalled, and the variations of the 3-metric near the boundary 
${\cal S}$ are decomposed. Then, in Section 3, the boundary condition 
above is introduced and the constraints are discussed. The fourth 
section is devoted to the investigation of the basic Hamiltonians and 
the 2-surface observables. In particular, we calculate its value in 
axi-symmetric spacetimes and the small and large sphere limits. 

Our notations and conventions are essentially those that used in 
[3,4,9]. In particular, we use the abstract index formalism, and the 
curvature is defined by $-R^a{}_{bcd}X^b:=(D_cD_d-D_dD_c)X^a$. Though 
primarily we are interested in the physical 3+1 dimensional case, the 
analysis will be done in $n+1$ dimensions, $n\geq2$, and the signature 
of the spacetime metric is $1-n$ (and hence the spatial metric is {\it 
negative} definite). Although here we consider only the vacuum case 
(with cosmological constant $\lambda$), in our formulae we retain the 
gravitational `coupling constant' $\kappa=8\pi G$. The analysis is 
based on certain formulae given explicitly in [3]. 

\bigskip
\bigskip

\ni
{\lbf 2 Variation of the constraint function}
\bigskip
\ni
Let $\Sigma$ be any smooth $n$ dimensional compact manifold with 
smooth $(n-1)$-boundary ${\cal S}:=\partial\Sigma$. Then the constraint 
function in the ADM phase space of the $n+1$ dimensional vacuum 
general relativity with cosmological constant $\lambda$, smeared by 
the function $N$ and vector field $N^a$ on $\Sigma$, is 

$$
C\bigl[N,N^a\bigr]:=-\int_\Sigma\Bigl\{{1\over2\kappa}\Bigl[R-2\lambda
+{4\kappa^2\over\vert h\vert}\bigl({1\over(n-1)}\tilde p^2-\tilde 
p_{ab}\tilde p^{ab}\bigr)\Bigr]N\sqrt{\vert h\vert}+2N^ch_{ca}D_b
\tilde p^{ab}\Bigr\}{\rm d}^nx. \eqno(2.1)
$$
Here the canonical variables are $h_{ab}$ and $\tilde p^{ab}$, $D_e$ 
is the Levi-Civita covariant derivative determined by $h_{ab}$ and 
$R$ is its curvature scalar. In spacetime this constraint function 
is just the integral $\int_\Sigma\xi^a(G_{ab}+\lambda g_{ab})t^b{\rm d}
\Sigma$, where $t^a$ is the future pointing unit timelike normal to 
$\Sigma$ in the spacetime, $\xi^a:=Nt^a+N^a$, and in the momentum 
phase space their vanishing for all $N$ and $N^a$ define the 
constraint surface $\Gamma$. 
The canonical momentum in terms of the Lagrange variables, i.e. the 
metric and the extrinsic curvature $\chi_{ab}$ of $\Sigma$ in the 
spacetime, is known to be $\tilde p^{ab}={1\over2\kappa}\sqrt{\vert h
\vert}(\chi^{ab}-\chi h^{ab})$. Here $\chi$ is the $h_{ab}$-trace of 
$\chi_{ab}$, the velocity of $h_{ab}$ is $\dot h_{ab}=2N\chi_{ab}+
\L_{\bf N}h_{ab}$ and $N$ and $N^a$ play the role of the lapse and 
the shift, respectively, in the spacetime. $\L_{\bf N}$ denotes the 
Lie derivative along $N^a$. 

Let $N(u)$, $N^a(u)$, $h_{ab}(u)$ and $\tilde p^{ab}(u)$, $u\in
(-\epsilon,\epsilon)$, be any smooth 1-parameter families of lapses, 
shifts, metrics and canonical momenta, respectively, and define the 
corresponding variation of any function of them, $F=F(N,N^a,h_{ab},
\tilde p^{ab})$, as $\delta F:=({\rm d}F(N(u),N^a(u),h_{ab}(u),\tilde 
p^{ab}(u))/{\rm d}u)\vert_{u=0}$. Then the corresponding variation of 
the constraint function $C[N,N^e]$, taken from [3], is 

$$\eqalign{
\delta C\bigl[N,N^e\bigr]=&C\bigl[\delta N,\delta N^e\bigr]+\int
  _\Sigma\Bigl({\delta C[N,N^e]\over\delta h_{ab}}\delta h_{ab}
  +{\delta C[N,N^e]\over\delta\tilde p^{ab}}\delta\tilde p^{ab}\Bigr)
  {\rm d}^nx+\cr
+{1\over2\kappa}\oint_{\partial\Sigma}&\Bigl\{N\bigl(h^{ab}v^e(D_e
  \delta h_{ab})-v^a(D^b\delta h_{ab})\bigr)+\bigl(v^aD^bN-h^{ab}v^e
  D_eN\bigr)\delta h_{ab}+\cr
&+{2\kappa\over\sqrt{\vert h\vert}}\bigl(2N^av_e\tilde p^{eb}-N^ev_e
  \tilde p^{ab}\bigr)\delta h_{ab}+4\kappa N_av_b{\delta\tilde p^{ab}
  \over\sqrt{\vert h\vert}}\Bigr\}{\rm d}{\cal S}.\cr} \eqno(2.2)
$$
Here ${\rm d}{\cal S}$ is the induced volume $n-1$-form on ${\cal S}$, 
$v^a$ is the {\it outward} pointing unit normal of ${\cal S}$ in 
$\Sigma$, and 

$$\eqalignno{
{\delta C[N,N^e]\over\delta h_{ab}}:=&{1\over2\kappa}\sqrt{\vert h
  \vert}\Bigl\{N\Bigl(R^{ab}-Rh^{ab}+2\lambda h^{ab}+{8\kappa^2\over
  \vert h\vert}\bigl(\tilde p^a{}_c\tilde p^{cb}-{1\over(n-1)}h_{cd}
  \tilde p^{cd}\tilde p^{ab}\bigr)\Bigr)+&(2.3.a)\cr
+&D^aD^bN-h^{ab}D_cD^cN\Bigr\}-\L_{\bf N}\tilde p^{ab}+{1\over4\kappa}
  Nh^{ab}\sqrt{\vert h\vert}\Bigl(R-2\lambda+{4\kappa^2\over\vert h\vert}
  \bigl({1\over(n-1)}\tilde p^2-\tilde p^{cd}\tilde p_{cd}\bigr)
  \Bigr),\cr
{\delta C[N,N^e]\over\delta\tilde p^{ab}}:=&{4\kappa\over\sqrt{\vert 
  h\vert}}N\Bigl(\tilde p^{ab}-{1\over(n-1)}\tilde p^{cd}h_{ca}h_{db}
  \Bigr)+\L_{\bf N}h_{ab}.&(2.3.b)\cr}
$$
Here $R_{ab}$ is the Ricci tensor of $D_e$. Thus $C[N,N^e]$ is 
functionally differentiable (in the strict sense of [14,15]) with 
respect to the canonical variables only if the boundary integral in 
(2.2) is vanishing, whenever the functional derivatives themselves are 
given by (2.3). Then the vacuum evolution equations with cosmological 
constant are precisely the canonical equations 

$$
\dot h_{ab}={\delta C\bigl[N,N^e\bigr]\over\delta\tilde p^{ab}}, 
\hskip 20pt
\dot{\tilde p}{}^{ab}=-{\delta C\bigl[N,N^e\bigr]\over\delta h_{ab}}, 
\eqno(2.4.a,b)
$$
provided the constraint equations $C[N,N^e]=0$ are satisfied. Our 
ultimate aim is to find appropriate boundary conditions on the 
canonical variables $(h_{ab},\tilde p{}^{ab})$ and an appropriate 
class of fields $N$, $N^a$ together with a boundary integral $B[N,
N^e]$ such that $C[N,N^e]+B[N,N^e]$ is functionally differentiable, 
and the boundary conditions on the canonical variables are compatible 
with the evolution equations. 

To find this boundary term and these conditions, it seems useful to 
split the variation of the metric $h_{ab}$ at the points of ${\cal S}$ 
with respect to the boundary. Thus let $\Pi^a_b:=\delta^a_b+v^av_b$, the 
$h_{ab}$-orthogonal projection to ${\cal S}$, and define the induced 
metric $q_{ab}:=h_{cd}\Pi^c_a\Pi^d_b$, the corresponding Levi-Civita 
covariant derivative $\delta_e$ and another derivative operator simply 
by $\Delta_e:=\Pi^f_eD_f$. The extrinsic curvature of ${\cal S}$ in 
$\Sigma$ will be defined by $\nu_{ab}:=\Pi^c_a\Pi^d_bD_cv_d$. At the 
points of ${\cal S}$ the splitting $h_{ab}=q_{ab}-v_av_b$ implies the 
variation $\delta h_{ab}=\delta q_{ab}-v_a\delta v_b-v_b\delta v_a$. 
Since $v_a$ is a normal 1-form of the submanifold ${\cal S}$, for 
any $X^a$ tangent to ${\cal S}$ one has $v_a(u)X^a=0$, implying that 
$\delta v_a\Pi^a_b=0$. Taking the $u$-derivative of $q_{ab}(u)v^a(u)=0$ 
we obtain that $\delta q_{ab}v^av^b=0$ and $\delta q_{ab}v^a\Pi^b_c=
-\delta v^aq_{ac}$, and taking the derivative of $v^a(u)v_a(u)=1$ we 
obtain $\delta v^av_a=-\delta v_av^a$. Thus, the various projections 
of the variation $\delta h_{ab}$ are 

$$
\delta h_{cd}\Pi^c_a\Pi^d_b=\delta q_{cd}\Pi^c_a\Pi^d_b, \hskip 20pt
\delta h_{cd}v^c\Pi^d_b=-\delta v^aq_{ab}, \hskip 20pt
\delta h_{cd}v^cv^d=2v^a\delta v_a=-2v_a\delta v^a. \eqno(2.5) 
$$
Therefore, the independent variations can be represented by $\delta 
q_{cd}\Pi^c_a\Pi^d_b$ and $\delta v^a$. 

\bigskip
\bigskip

\ni
{\lbf 3 The quasi-local constraint algebra}
\bigskip
\ni
In this section we determine the boundary conditions under which the 
constraint functions are functional differentiable with respect to 
the canonical variables. We will see that, as a bonus, this already 
ensures that they form a Poisson algebra too. (In the asymptotically 
flat case it has been demonstrated that in vacuum general relativity 
this differentiability implies the Poisson algebra structure [2]. 
Similar result has been proven in a more general classical field 
theory context in [14]: functional differentiability of functions 
together with the requirement that the corresponding Hamiltonian 
vector fields preserve the boundary conditions also implies the Poisson 
algebra structure.) Thus first let us determine the condition of 
the functional differentiability of $C[N,N^a]$. To do this, we 
decompose the boundary integral in (2.2) with respect to ${\cal S}$. 
Clearly, $C[N,N^a]$ is functionally differentiable with respect to 
$N$ and $N^a$, independently of the boundary conditions at ${\cal 
S}$. A tedious but straightforward calculation yields that the 
vanishing of the boundary integral in (2.2) is just the condition 

$$\eqalign{
0=\oint_{\cal S}\Bigl({1\over2\kappa}&N\bigl(v^a(D^b\delta h_{ab})-
 h^{ab}v^e(D_e\delta h_{ab})\bigr)-{1\over2\kappa}v^a\delta h_{ab}
 q^{bc}\Delta_cN+{1\over2\kappa}v^e(D_eN)q^{ab}\delta h_{ab}-\cr
-2v_e&{\tilde p{}^{ef}
 \over\sqrt{\vert h\vert}}N^d\Pi^a_f\Pi^b_d\delta h_{ab}
+v_eN^e\bigl({\tilde p{}^{ab}\over\sqrt{\vert h\vert}}\delta h_{ab}+
 2v_e{\tilde p{}^{ea}\over\sqrt{\vert h\vert}}\Pi^b_a\delta h_{bc}v^c
 -2v_ev_f{\tilde p{}^{ef}\over\sqrt{\vert h\vert}}v^av^b\delta h_{ab}
 \bigr)+\cr
+2v_ev_f&{\tilde p{}^{ef}\over\sqrt{\vert h\vert}}v^a\delta h_{ab}
 \Pi^b_cN^c\Bigr){\rm d}{\cal S}.\cr} \eqno(3.1)
$$
Taking into account that the variation of the induced volume 
$(n-1)$-form on the boundary is $\delta\varepsilon_{a_1...a_{n-1}}=
{1\over2}q^{cd}\delta q_{cd}\varepsilon_{a_1...a_{n-1}}={1\over2}
q^{cd}\delta h_{cd}\varepsilon_{a_1...a_{n-1}}$, the boundary 
conditions $N\vert_{\cal S}=0$, $N^a\vert_{\cal S}=0$ and $\varepsilon
_{a_1...a_{n-1}}={\rm fixed}$ ensure the functional differentiability 
of the constraint functions $C[N,N^a]$ with respect to $h_{ab}$ and 
$\tilde p{}^{ab}$. 
Since the last term of the integrand in (3.1) is proportional to 
$2\kappa v_av_b\tilde p{}^{ab}=\sqrt{\vert h\vert}q^{ab}\chi_{ab}$, 
which is not zero in general, the boundary condition $N^a\vert_{\cal 
S}=0$ cannot be weakened to $v_aN^a\vert_{\cal S}=0$ even if the 
induced metric $q_{ab}$ on ${\cal S}$ (rather than only the 
corresponding volume $(n-1)$-form) is kept fixed. 
On the other hand, because of the fourth term in (3.1), $N\vert_{\cal 
S}=0$ and $N^a\vert_{\cal S}=0$ in themselves are not enough to 
ensure the functional differentiability with respect to $h_{ab}$. 

These boundary conditions are preserved by the evolution equations. 
Indeed, since the only condition that we imposed on the canonical 
variables is $\delta\varepsilon_{a_1...a_{n-1}}=0$, we should consider 
only (2.4.a), the evolution equation for the metric $h_{ab}$. By 
$N\vert_{\cal S}=0$ this yields on the boundary that $\dot h_{ab}\vert
_{\cal S}=2D_{(a}N_{b)}$, and hence, by (2.5), $q^{ab}\dot q_{ab}=q
^{ab}\dot h_{ab}=2q^{ab}D_aN_b=2\Delta_aN^a=0$, where in the last step 
we used $N^a\vert_{\cal S}=0$. Therefore, {\it the evolution equations 
preserve the boundary conditions}. Geometrically $N\vert_{\cal S}=0$, 
$N^a\vert_{\cal S}=0$ correspond to an evolution vector field $\xi^a
=t^aN+N^a$ in the spacetime that is vanishing on ${\cal S}$; i.e. the 
corresponding diffeomorphism leaves ${\cal S}$ fixed {\it pointwise}. 
The one parameter family of diffeomorphisms generated by such a 
$\xi^a$ maps $\Sigma$ into a family $\Sigma_t$ of Cauchy surfaces for 
the {\it same} globally hyperbolic domain $D(\Sigma)$ with the same 
boundary $\partial\Sigma_t={\cal S}$, i.e. such a $\xi^a$ is precisely 
a vector field that we would intuitively consider to be the generator 
of a gauge motion in the {\it spacetime}. 

By the functional differentiability of the constraint functions (with 
vanishing smearing fields $N$ and $N^a$ on ${\cal S}$) we can take the 
Poisson bracket of any two constraint functions $C[N,N^a]$ and $C[\bar 
N,\bar N^a]$. These brackets, keeping all the boundary terms, have 
already been calculated [3]. They are 

$$\eqalignno{
\Bigl\{C\bigl[0,N^a\bigr],C\bigl[0,\bar N^a\bigr]\Bigr\}=&-C\bigl[0,
 [N,\bar N]^a\bigr]+\cr
+\int_\Sigma D_e\Bigl(&N^e\tilde p{}^{ab}\L_{\bar N}h_{ab}-\bar N^e
 \tilde p{}^{ab}\L_{N}h_{ab}-2\tilde p{}^{ef}h_{fa}[N,\bar N]^a\Bigr)
 {\rm d}^nx, &(3.2.a)\cr
\Bigl\{C\bigl[0,N^a\bigr],C\bigl[\bar N,0\bigr]\Bigr\}=&-C\bigl[N^e
 D_e\bar N,0\bigr]+\cr
+{1\over\kappa}\int_\Sigma D_e\Bigl(&\bar N\bigl(R^e{}_f-{1\over2}R
 \delta^e_f\bigr)N^f+\lambda\bar NN^e+{2\kappa^2\over\vert h\vert}
 \bar NN^e\bigl(\tilde p{}^{ab}\tilde p_{ab}-{1\over n-1}\tilde p
 {}^2\bigr)+\cr
+&\bigl(\Delta_fN^e\bigr)\bigl(\Delta^f\bar N\bigr)-\bigl(D^e\bar N
 \bigr)\bigl(\Delta_fN^f\bigr)\Bigr)\sqrt{\vert h\vert}{\rm d}^nx,
 &(3.2.b)\cr
\Bigl\{C\bigl[N,0\bigr],C\bigl[\bar N,0\bigr]\Bigr\}=&C\bigl[0,
 ND^a\bar N-\bar ND^aN\bigr]+
2\int_\Sigma D_e\Bigl(N\tilde p{}^{ef}D_f\bar N-\bar N\tilde p{}
 ^{ef}D_fN\Bigr){\rm d}^nx. &(3.2.c)\cr}
$$
However, by the vanishing of the smearing fields on ${\cal S}$ all 
the boundary terms in (3.2) are vanishing, and the Lie product can 
be summarized as 

$$
\Bigl\{C\bigl[N,N^a\bigr],C\bigl[\bar N,\bar N^a\bigr]\Bigr\}=C\bigl[
\bar N^eD_eN-N^eD_e\bar N,ND^a\bar N-\bar ND^aN-[N,\bar N]^a\bigr]. 
\eqno(3.3)
$$
Furthermore, the new smearing fields $\bar N^eD_eN-N^eD_e\bar N$ and 
$ND^a\bar N-\bar ND^aN-[N,\bar N]^a$ are also vanishing on the 
boundary ${\cal S}$. Therefore, {\it the constraint functions with 
vanishing smearing fields on ${\cal S}$ close to a Poisson algebra 
${\cal C}$}, the so-called quasi-local constraint algebra, provided 
the induced volume $(n-1)$-form $\varepsilon_{a_1...a_{n-1}}$ is 
fixed on ${\cal S}$. Clearly, this Lie algebra is isomorphic to that 
appearing in the asymptotically flat case [2-4]. 

The boundary condition yields the split of the quasi-local phase 
space $T^*{\cal Q}(\Sigma):=\{(h_{ab},\tilde p{}^{ab})\}$ into the 
disjoint union of sectors $T^*{\cal Q}(\Sigma,\varepsilon_{a_1...a
_{n-1}})$, labelled by the volume $(n-1)$-from $\varepsilon_{a_1...
a_{n-1}}$ on ${\cal S}$: The constraint functions are differentiable 
in the directions tangent to these sectors and form the familiar 
Poisson algebra, and the evolution equations with lapse and shift 
vanishing on ${\cal S}$ also preserve this sector--structure. 

\bigskip
\bigskip

\ni
{\lbf 4 The basic Hamiltonian}
\medskip

\ni
{\bf 4.1 The boundary conditions}
\smallskip
\ni
Starting with the naive quasi-local Lagrange phase space $T{\cal Q}
(\Sigma):=\{(h_{ab},\dot h_{ab})\}$ and the traditional Lagrangian 
$L:T{\cal Q}(\Sigma)\rightarrow{\bf R}$, given explicitly by $L:=
{1\over2\kappa}\int_\Sigma N(R-2\lambda+\chi^{ab}\chi_{ab}-\chi^2)
\sqrt{\vert h\vert}{\rm d}^nx$, the basic Hamiltonian $H_0[N,N^a]:=
\int_\Sigma\tilde p{}^{ab}\dot h_{ab}{\rm d}^nx-L$ on $T^*{\cal Q}
(\Sigma)$ takes the form 

$$
H_0\bigl[N,N^e\bigr]=C\bigl[N,N^e\bigr]+\int_\Sigma2D_a\Bigl(\tilde 
p{}^{ab}h_{bc}N^c\Bigr){\rm d}^nx. \eqno(4.1)
$$
Its total variation is 

$$\eqalign{
\delta H_0\bigl[N,N^e\bigr]=&C\bigl[\delta N,\delta N^e\bigr]+\int
  _\Sigma\Bigl({\delta C[N,N^e]\over\delta h_{ab}}\delta h_{ab}+
  {\delta C[N,N^e]\over\delta\tilde p^{ab}}\delta\tilde p^{ab}\Bigr)
  {\rm d}^nx+\cr
+{1\over2\kappa}\oint_{\partial\Sigma}\Bigl\{&N\bigl(h^{ab}v^e(D_e
  \delta h_{ab})-v^a(D^b\delta h_{ab})\bigr)+\bigl(v^aD^bN-h^{ab}
  v^eD_eN\bigr)\delta h_{ab}-\cr
&-{2\kappa\over\sqrt{\vert h\vert}}\bigl(v_eN^e\tilde p^{ab}\delta h
  _{ab}+2v_e\tilde p^{ea}h_{ab}\delta N^b\bigr)\Bigr\}{\rm d}{\cal S}.
 \cr} \eqno(4.2)
$$
Thus $H_0[N,N^e]$ is functionally differentiable with respect to $N$ 
and the canonical momentum $\tilde p{}^{ab}$, independently of the 
boundary conditions at ${\cal S}$. 

The condition of the functional differentiability of $H_0[N,N^e]$ 
with respect to $h_{ab}$ is the vanishing of the boundary term in 
(4.2) involving $\delta h_{ab}$, provided the variations $\delta h
_{ab}$ and $\delta N^a$ are independent. We decompose its integrand 
with respect to the boundary $(n-1)$-surface, using spacetime 
quantities as well. In particular, if $t^a$ is the future pointing 
unit timelike normal to $\Sigma$ in spacetime and $A_e:=v_a\Delta
_et^a:=v_a\Pi^f_e\nabla_ft^a$, the connection 1-form on the normal 
bundle of ${\cal S}$, where now $\Pi^a_b:=\delta^a_b+v^av_b-t^at_b$ 
is the $g_{ab}$-orthogonal projection to ${\cal S}$ (for the details 
see [9] and references therein), then a lengthy but direct calculation 
gives that it is 

$$\eqalign{
0=\oint_{\cal S}\Bigl\{-Nv^e\Bigl(&D_e\delta h_{ab}\Bigr)q^{ab}+
 \delta h_{ab}v^av^b\Bigl(-N(\Delta_ev^e)+v_fN^f(\Delta_et^e)\Bigr)+
 \delta h_{ab}v^aq^{bc}\Bigl(-2\Delta_cN-2A_cv_eN^e\Bigr)+\cr
+\delta h_{ab}q^{ac}q^{bd}\Bigl(&q_{cd}v^e(D_eN)-N(\Delta_cv_d)+
 v_eN^e(\Delta_ct_d)-q_{cd}v_eN^e(\Delta_ft^f)+q_{cd}v_eN^ev^f(D_f
 t_g)v^g\Bigr)\Bigr\}{\rm d}{\cal S}.\cr}\eqno(4.3)
$$
The simplest way to make the first term vanishing is the condition 
that $N$ be vanishing on ${\cal S}$, whenever $v_eN^e\vert_{\cal S}
=0$ and $\varepsilon_{a_1...a_{n-1}}={\rm fixed}$ already ensure the 
functional differentiability of $H_0[N,N^e]$ with respect to $h_{ab}$. 
Note that this condition is weaker than that we had for the constraint 
functions, because we should require only that $N^a$ be tangent to 
${\cal S}$ rather than vanishing on ${\cal S}$. If we want $(n-1)
+2$-covariant conditions for $N$ and $N^a$ at ${\cal S}$, then by 
$N\vert_{\cal S}=0$ we must impose $v_aN^a\vert_{\cal S}=0$ too. 
Indeed, if we do not want to prefer any timelike normal to ${\cal S}$, 
then $N$ and $v_aN^a$ must be treated on an equal footing, because 
they are the two components of $\xi^a=t^aN+N^a$ orthogonal to ${\cal 
S}$. On the other hand, in the absence of additional conditions we 
loose the functional differentiability with respect to $N^a$. 

By $N\vert_{\cal S}=0$ the evolution equation for the metric gives 
$q^{ab}\dot q_{ab}=q^{ab}\dot h_{ab}=2q^{ab}D_aN_b=2\Delta_aN^a=
\delta_aN^a$, where in the last step we used $v_aN^a\vert_{\cal S}
=0$. Therefore, in addition, {\it we must require that $N^a$ on 
${\cal S}$ be divergence-free with respect to the intrinsic geometry 
of ${\cal S}$ as well}, otherwise the evolution equations do not 
preserve the boundary condition $\varepsilon_{a_1...a_{n-1}}={\rm 
fixed}$. At first sight the requirement that $N^a$ on ${\cal S}$ be 
$\delta_e$-divergence-free yields that the variation of the metric 
on ${\cal S}$ produces a variation of $N^a$ on ${\cal S}$, and hence 
these variations on ${\cal S}$ are not quite independent. However, 
by $\delta(\delta_aN^a)=N^e\delta_e({1\over2}q^{ab}\delta q_{ab})+
\delta_a(\delta N^a)$ and the boundary condition $q^{ab}\delta q
_{ab}=0$ the variation of the metric alone does not yield any variation 
of $\delta_aN^a$. In other words, if $N^a$ is any shift such that 
$v_aN^a\vert_{\cal S}=0$ and $N^a$ is $\delta_e$-divergence-free, 
then it will be divergence-free with respect to the connection coming 
from any 1-parameter family $q_{ab}(u)$ of metrics provided the 
volume $(n-1)$-form is kept fixed. Geometrically, $N\vert_{\cal S}=0$ 
and $v_aN^a\vert_{\cal S}=0$ correspond to an evolution vector field 
$\xi^a$ in the spacetime which is tangent to ${\cal S}$, and hence, 
by $\delta_aN^a\vert_{\cal S}=0$, it generates a volume preserving 
diffeomorphism of ${\cal S}$ to itself. 

\bigskip
\ni
{\bf 4.2 The algebra of the basic Hamiltonians and 2-surface 
observables}
\smallskip
\ni
Since the formal variational derivatives of the constraint functions 
and of the basic Hamiltonians are the same, the Poisson bracket of 
two basic Hamiltonians, $H_0[N,N^a]$ and $H_0[\bar N,\bar N^a]$, can 
be calculated easily by (3.2). By the boundary conditions $v_aN^a
\vert_{\cal S}=v_a\bar N^a\vert_{\cal S}=0$ the boundary term in the 
Poisson bracket $\{H_0[0,N^a],H_0[0,\bar N^a]\}$ is vanishing, and 
there is no boundary term at all in the Poisson bracket $\{H_0[N,0],
H_0[\bar N,0]\}$. On the other hand, the boundary term in the Poisson 
bracket $\{H_0[0,N^a],H_0[\bar N,0]\}$ is vanishing only if we use 
$\delta_aN^a\vert_{\cal S}=0$ too. This gives an additional 
justification of the condition $\delta_aN^a\vert_{\cal S}=0$. 
Then the Lie product of the basic Hamiltonians can be summarized as 

$$
\Bigl\{H_0\bigl[N,N^a\bigr],H_0\bigl[\bar N,\bar N^a\bigr]\Bigr\}=H_0
\bigl[\bar N^eD_eN-N^eD_e\bar N,ND^a\bar N-\bar ND^aN-[N,\bar N]^a
\bigr]. \eqno(4.4)
$$
Furthermore, if $N^a$ and $\bar N^a$ are any two shifts which are 
tangent to ${\cal S}$ and $\delta_a$-divergence-free on ${\cal S}$, 
then their Lie bracket $[N,\bar N]^a$ is also tangent to ${\cal S}$ 
and $\delta_a$-divergence-free on ${\cal S}$. Hence the new lapse 
$\bar N^eD_eN-N^eD_e\bar N$ and the new shift $ND^a\bar N-\bar ND^a
N-[N,\bar N]^a$ also satisfy the boundary conditions. Therefore, 
{\it the basic Hamiltonians parameterized by lapses and shifts 
satisfying $N\vert_{\cal S}=0$, $v_aN^a\vert_{\cal S}=0$ and $\delta
_aN^a\vert_{\cal S}=0$ form a Poisson algebra ${\cal H}_0$}. 

The value of the basic Hamiltonian on the constraint surface is 

$$
O\bigl[N^a\bigr]:=H_0\bigl[N,N^a\bigr]\vert_\Gamma=-{1\over\kappa}
\oint_{\cal S}N^aA_a{\rm d}{\cal S}. \eqno(4.5)
$$
Though $A_a$ is not a gauge invariant object (namely, as we already 
mentioned, this is a connection 1-form in the normal bundle of ${\cal 
S}$ in the spacetime, and under an $SO(1,1)$ boost gauge transformation 
of the two normals, $(t^a,v^a)\mapsto(t^a\cosh(w)+v^a\sinh(w),v^a\cosh
(w)+t^a\sinh(w))$, it transforms as a vector potential), by $\delta_a
N^a\vert_{\cal S}=0$ the integral $O[N^a]$ is indeed boost gauge 
invariant. This is the third justification of the condition $\delta_a
N^a\vert_{\cal S}=0$. 
Clearly, the constraint functions form an ideal in the algebra of 
the basic Hamiltonians, ${\cal C}\subset{\cal H}_0$, and the quotient 
${\cal H}_0/{\cal C}$ can be parameterized by the value $O[N^a]$. 
By (4.4) this $O[N^a]$ defines a Lie algebra anti-homomorphism of the 
Lie algebra of the divergence-free vector fields on ${\cal S}$ into 
${\cal H}_0/{\cal C}$: 
in fact, let $N^a$, $N'^a$ and $\bar N^a$, $\bar N'^a$ be shift vectors 
such that they are tangent to ${\cal S}$ and $\delta_e$-divergence-free 
on ${\cal S}$, furthermore $N^a\vert_{\cal S}=N'^a\vert_{\cal S}$ and 
$\bar N^a\vert_{\cal S}=\bar N'^a\vert_{\cal S}$. Then $O[N^a]=
O[N'^a]$ and $\{H_0[0,N^a],H_0[0,\bar N^a]\}\vert_\Gamma=\{H_0[0,N'^a],
H_0[0,\bar N'^a]\}\vert_\Gamma$, i.e. both $O[N^a]$ and the Poisson 
bracket $\{H_0[0,N^a],H_0[0,\bar N^a]\}$ evaluated on the constraint 
surface depend only on the restriction of the shifts to ${\cal S}$, and 
independent of their part inside $\Sigma$. Hence the Poisson bracket 
$\{O[N^a],O[\bar N^a]\}:=\{H_0[0,N^a],H_0[0,\bar N^a]\}\vert_\Gamma$ of 
$O[N^a]$ and $O[\bar N^a]$ is well defined and, by (4.4), it is 
$\{O[N^a],O[\bar N^a]\}=-O[[N,\bar N]^a]$. 

It might be worth noting that the $\delta_e$-divergence free vector 
fields on ${\cal S}$ can be given explicitly by using the Hodge 
decomposition (see e.g. [16]): if $N^a$ is divergence free, then it 
necessarily has the form $\delta_bN^{ab}+\ast\omega^a$, where $N^{ab}=
N^{[ab]}$ is an arbitrary bi-vector and $\ast\omega^a:={1\over(n-2)!}
\varepsilon^{aa_1...a_{n-2}}\omega_{a_1...a_{n-2}}$ denotes the Hodge 
dual of a harmonic $(n-2)$-form $\omega_{a_1...a_{n-2}}$. The latter 
is an arbitrary linear combination of finitely many linearly independent 
harmonic forms $\omega_{a_1...a_{n-2}}^\alpha$, $\alpha=1,...,b$, where 
$b:=\dim H^{n-2}({\cal S})$, the $(n-2)$th Betti number of ${\cal S}$. 
In terms of these the observable (4.5) takes the form $-{1\over
\kappa}\oint_{\cal S}(N^{ab}\delta_aA_b+\ast\omega^aA_a){\rm d}{\cal 
S}$. In the physically important special case $n=3$ the bi-vector can 
always be written as $\varepsilon^{ab}\nu$ with an arbitrary real 
function $\nu$, and the Betti number is $b=2g$, twice the genus of 
${\cal S}$. 

Formally, $O[N^a]$ is just the observable $O_M[N^a]$ of Balachandran, 
Chandar and Momen [5,6] (see also [7,8]). However, the present 
boundary conditions for the canonical variables are definitely weaker 
than those of them: they kept fixed the whole $n$-metric $h_{ab}$ on 
${\cal S}$. On the other hand, without the extra condition $\delta_a
N^a\vert_{\cal S}=0$, the observable $O_M[N^a]$ is {\it not} 
boost-gauge invariant. In addition, this extra condition on $N^a$ 
ensures that the evolution equations preserve the weaker boundary 
conditions. Without this the evolution equations would preserve neither 
the boundary conditions of [5,6,8] nor the present, weaker ones. 
Similarly, the `natural' boundary condition that the induced $(n-
1)$-metric 
$q_{ab}$ is fixed is preserved by the evolution equation (2.4.a) only 
if $N^a$ is vanishing on ${\cal S}$ or if $({\cal S},q_{ab})$ admits 
$N^a$ as a Killing vector. It could be interesting to note that the 
quasi-local quantity $L(N^a)$ of Yoon [17], obtained by following an 
$(n-1)+2$ analysis of the vacuum Einstein equations, as well as the 
`(generalized) angular momentum' of Brown and York [18], of Liu and 
Yau [19], and of Ashtekar and Krishnan [20] are just the observable 
$O[N^a]$ provided $N^a$ on ${\cal S}$ is restricted to be tangent to 
${\cal S}$ and $\delta_e$-divergence-free on ${\cal S}$. Another (and 
quite obvious) observable is the surface integral of any integrable 
`test' function $f$ on ${\cal S}$: $A[f]:=\oint_{\cal S}f{\rm d}{\cal 
S}$. 

In [5,6] a further `observable' $O_H[T]$ was introduced, where $T$ is 
the (not necessarily vanishing) {\it constant} value of the lapse on 
${\cal S}$, and this was interpreted as some (not renormalized) form 
of energy. However, it depends on the choice for a preferred timelike 
normal to ${\cal S}$ too; i.e. {\it not} boost gauge invariant.

\bigskip
\ni
{\bf 4.3 The various limits of the 2-surface observable}
\smallskip
\ni
To clarify the meaning of the observable $O[N^a]$ it seems natural 
to consider various special 3+1 dimensional spacetimes and limits, 
such as axi-symmetric spacetimes, and the small and large sphere 
limits. 

\medskip
\ni
{$\bullet$ {\it Axi-symmetric spacetimes}

\ni
Let the spacetime be axi-symmetric with Killing vector $K^a$. Then 
the angular momentum is usually defined by the 2-surface integral 
of the Komar superpotential built from $K^a$, and the value of this 
integral is well known to be invariant with respect to the continuous 
deformations of the 2-surface through {\it vacuum} regions (see e.g. 
[15]). To be able to compare the Komar expression and the observable, 
let us fix the 2-surface ${\cal S}$ and a foliation $\Sigma_t$ of an 
open neighbourhood of ${\cal S}$ by smooth spacelike hypersurfaces 
such that ${\cal S}$ is lying in one leaf, e.g. in $\Sigma_0$, and 
let $v^a$ denote the outward pointing unit normal of ${\cal S}$ in 
$\Sigma_0$. (This foliation should not be confused with the foliation 
of the globally hyperbolic domain whose `edge' is the 2-surface 
${\cal S}$: the former foliates an open neighbourhood of ${\cal S}$, 
whilst the latter collapses just on ${\cal S}$.) Let $t^a$ be the 
future pointing unit normal of the leaves of the foliation, $P^a_b:=
\delta^a_b-t^at_b$ the orthogonal projection to the leaves and let 
$M$ be the lapse function of the foliation. Let us choose a shift 
vector $M^a$ as well, i.e. specify an `evolution vector field' $\xi
^a:=Mt^a+M^a$. Then let $K^a=:Nt^a+N^a$ define the 3+1 pieces of the 
Killing field $K^a$ with respect to the foliation. Then the 
time--space projection of the Killing operator acting on $K_a$, taken 
from [3,4], is 

$$
2MP^c_at^d\nabla_{(c}K_{d)}=\bigl(\L_\xi N_b\bigr)P^b_a-\bigl(\L
_MN_b\bigr)P^b_a+MD_aN-ND_aM-2M\chi_{ab}N^b, \eqno(4.6)
$$
where $h_{ab}=g_{ab}-t_at_b$ is the induced metric on and $\chi
_{ab}$ is the extrinsic curvature of the leaves. Using this, Komar's 
expression (normalized to get the correct value for the angular 
momentum in Kerr spacetime, see [21]) can be written as 

$$\eqalign{
{\tt I}_{\cal S}\bigl[K^a\bigr]:&={1\over2\kappa}\oint_{\cal S}\nabla
 ^{[a}K^{b]}{1\over2}\varepsilon_{abcd}={1\over\kappa}\oint_{\cal S}
 v^aP^c_at^d\bigl(\nabla_{[c}K_{d]}\bigr){\rm d}{\cal S}=\cr
&={1\over\kappa}\oint_{\cal S}v^a\Bigl(-{1\over2M}\bigl(\L_\xi N_b
 \bigr)P^b_a+{1\over2M}\bigl(\L_MN_b\bigr)P^b_a+{1\over2M}D_a\bigl(NM
 \bigr)\Bigr){\rm d}{\cal S}=\cr
&={1\over\kappa}\oint_{\cal S}\Bigl(v^aD_aN-v^a\chi_{ab}N^b-v^aP^c_a
 t^d\nabla_{(c}K_{d)}\Bigr){\rm d}{\cal S}.\cr} \eqno(4.7)
$$
Thus if the 2-surface ${\cal S}$ is chosen to be axi-symmetric (i.e. 
if $K^a$ is tangent to ${\cal S}$ on ${\cal S}$) and $K^a$ is tangent 
to $\Sigma_0$, then by $K^a=N^a$ the first term of the integrand is 
vanishing, the second term is $-N^aA_a$, and the third term is also 
zero because $K^a$ is a Killing vector. Hence, in the special boost 
gauge defined by the hypersurface $\Sigma_0$ containing the integral 
curves of $K^a$, the observable $O[N^a]$ coincides with the Komar 
integral. Since, however, $O[N^a]$ is boost gauge invariant, we 
obtained that {\it the observable $O[N^a]$ for the Killing vector of 
axi-symmetry $N^a$ and for the axi-symmetric 2-surface ${\cal S}$ 
coincides with the Komar integral ${\tt I}_{\cal S}[N^a]$}. 
Since ${\tt I}_{\cal S}[K^a]$} is invariant with respect to continuous 
deformations of ${\cal S}$ through vacuum regions, and in the definition 
of the Komar integral $K^a$ is not required to be tangent to ${\cal S}$, 
the 2-surface is not required to be axi-symmetric. On the other hand, 
the observable $O[N^a]$ is well defined only for vector fields $N^a$ 
tangent to the 2-surface, and hence ${\cal S}$ should be required to be 
axi-symmetric. Thus for axi-symmetric surfaces the observable $O[N^a]$ 
reproduces Komar's angular momentum, but for non-axi-symmetric surfaces 
in an axi-symmetric spacetime, whenever Komar's expression can still be 
calculated, $O[N^a]$ is not even well defined. 

\medskip
\ni
{$\bullet$ {\it The small sphere limit}

\ni
To calculate $O[N^a]$ for small spheres ${\cal S}_r$ of radius $r$ 
about a point $p\in M$ defined by the future pointing unit timelike 
vector $t^a$ at $p$ (for the standard definitions of all these 
limits see e.g. [9] and references therein), it seems more convenient 
to use the expression of $N^a$ obtained form the application of the 
Hodge decomposition. Since no non-trivial harmonic form exists on 
spheres, we can write $N^a=\varepsilon^{ab}\delta_b\nu$ and $\nu$ is 
an arbitrary real function on ${\cal S}_r$. Since the field strength 
$-\varepsilon^{ab}\delta_aA_b$ is half the imaginary part of the 
complex Gauss curvature of ${\cal S}_r$ given in the well known GHP 
formalism by $K=-\psi_2-\rho\rho'+\sigma\sigma'+\phi_{11}+\Lambda$, 
the observable (4.3) takes the form 

$$
O\bigl[N^a\bigr]={{\rm i}\over\kappa}\oint_{{\cal S}_r}\nu\Bigl(
\psi_2-\bar\psi_{2'}-\sigma\sigma'+\bar\sigma\bar\sigma'\Bigr){\rm 
d}{\cal S}_r. \eqno(4.8)
$$
Expanding the Weyl spinor component as $\psi_2=\psi_2^{(0)}+r\psi_2
^{(1)}+r^2\psi_2^{(2)}+...$ and substituting the solution of the Ricci 
identities for $\sigma$ and $\sigma'$ and the expression of ${\rm d}
{\cal S}_r$ from [22] to (4.8), we obtain ${{\rm i}\over\kappa}\oint
_{{\cal S}_1}\nu(r^2[\psi_2^{(0)}-\bar\psi_{2'}^{(0)}]+r^3[\psi_2^{(1)}
-\bar\psi_{2'}^{(1)}]+r^4([\psi_2^{(2)}-\bar\psi_{2'}^{(2)}]-{1\over3}
\psi_{00}^{(0)}[\psi_2^{(0)}-\bar\psi_{2'}^{(0)}]+{2\over9}\phi_{20}
^{(0)}\psi_0^{(0)}-{2\over9}\phi_{02}^{(0)}\bar\psi_{0'}^{(0)})+O(r
^5)){\rm d}{\cal S}_1$. (${\rm d}{\cal S}_1$ is, of course, the unit 
sphere area element.) To have a definite expression, we must specify 
the function $\nu$ by hand. Since $O[N^a]$ is usually expected to be 
something similar to spatial angular momentum, let us suppose that 
$N^a$ is a linear combination of the three independent approximate 
spatial rotation Killing vectors in a neighbourhood of $p$ that vanish 
at $p$ and tangent to ${\cal S}_r$: 

$$
N^a={2\sqrt{2}r\over1+\zeta\bar\zeta}\Bigl(\bar m^a\bigl(M_{00}\zeta
^2+2M_{01}\zeta+M_{11}\bigr)+m^a\bigl(\bar M_{0'0'}\bar\zeta^2+2\bar 
M_{0'1'}\bar\zeta+\bar M_{1'1'}\bigr)\Bigr) +O\bigl(r^2\bigr).
\eqno(4.9)
$$
Here $M_{\uA\uB}=M_{(\uA\uB)}=(M_{00},M_{01},M_{11})$ are complex 
constants satisfying $\bar M_{1'1'}=M_{00}$ and $M_{01}$ is purely 
imaginary. (In Minkowski spacetime the leading order part of $N_e$ 
is precisely the $2(M_{\uA\uB}K^{\uA\uB}_f+\bar M_{{\uA}'{\uB}'}
\bar K^{{\uA}'{\uB}'}_f)\Pi^f_e$ combination of the anti-self-dual 
boost-rotation Killing 1-forms $K^{\uA\uB}_e$ that vanish at $p$. 
For the details see [22].) Then the corresponding function $\nu$ is 
$4{\rm i}r^2(1+\zeta\bar\zeta)^{-1}(M_{00}\zeta+2M_{01}-M_{11}\bar
\zeta)+O(r^3)$. Substituting this into the general $r^4$ accurate 
approximate formula above we obtain that $O[N^a]$ is vanishing in 
the $r^4$ order, and in non-vacuum the first non-vanishing order is 
$r^5$. In vacuum $O[N^a]$ is vanishing in all orders up to (and 
including) $r^6$. Since here we considered only approximate {\it 
rotation} (but not boost) Killing fields, this result is compatible 
with the expectations of [9,22]: Although in general non-vacuum 
spacetime the leading term in the small sphere expression of any 
reasonable angular momentum expression must be of order $r^4$ and 
in vacuum it must be of order $r^6$, but these correspond to the 
centre-of-mass part of the relativistic angular momentum. The 
rotation part is expected to be only of order $r^5$ and $r^7$, 
respectively. 

\medskip
\ni
{$\bullet$ {\it Large spheres near the future null infinity}

\ni
If ${\cal S}_r$ is a large sphere of radius $r$ near the future null 
infinity (see e.g. [23]), then we can write $O[N^a]$ into the form 
(4.8). Taking into account the asymptotic form of the Weyl spinor 
component and the shears given in [23], and writing the function 
$\nu$ as $\nu=r^2\nu^{(2)}+r\nu^{(1)}+\nu^{(0)}+O(r^{-1})$, (4.8) 
takes the form 

$$\eqalignno{
O\bigl[N^a\bigr]={{\rm i}\over\kappa}\oint_{{\cal S}_1}\Bigl\{r&\nu
 ^{(2)}\bigl({}_0{\edth}'^2\sigma^0-{}_0{\edth}^2\bar\sigma^0\bigr)
 +&(4.10)\cr
+&\nu^{(2)}\Bigl({}_0{\edth}\bigl(\bar\psi_{1'}^0+\bar\sigma^0\,{}_0
 {\edth}'\sigma^0\bigr)-{}_0{\edth}'\bigl(\psi_1^0+\sigma^0\,{}_0
 {\edth}\bar\sigma^0\bigr)\Bigr)+\nu^{(1)}\bigl({}_0{\edth}'^2\sigma^0
 -{}_0{\edth}^2\bar\sigma^0\bigr)\Bigr\}{\rm d}{\cal S}_1+O\bigl(r
 ^{-1}\bigr),\cr}
$$
where ${}_0{\edth}$ is the standard edth operator on the metric unit 
sphere. $O[N^a]$ has finite $r\rightarrow\infty$ limit precisely when 
$\nu^{(2)}\in{\rm ker}{}_0{\edth}^2\cap{\rm ker}{}_0{\edth}'^2$, or, 
explicitly, if $\nu^{(2)}=T^{\ua}t_{\ua}$ where $T^{\ua}$ are 
arbitrary real numbers, ${\ua}=0,...,3$, and $t_0:=1$, $t_1:=-(\bar
\zeta+\zeta)(1+\zeta\bar\zeta)^{-1}$, $t_2:=-{\rm i}(\bar\zeta-\zeta)
(1+\zeta\bar\zeta)^{-1}$, and $t_3:=-(\zeta\bar\zeta-1)(1+\zeta\bar
\zeta)^{-1}$. (These are precisely the components of the independent 
BMS translations [24].) Then we have ${}_0{\edth}\nu^{(2)}=-2^{-{1
\over2}}(1+\zeta\bar\zeta)^{-1}T^{\bi}\xi_{\bi}$ and ${}_0{\edth}'
\nu^{(2)}=-2^{-{1\over2}}(1+\zeta\bar\zeta)^{-1}T^{\bi}\bar\xi_{\bi}$, 
where ${\bi}=1,2,3$, and $\xi_1:=1-\zeta^2$, $\xi_2:={\rm i}(1+\zeta
^2)$ and $\xi_3:=2\zeta$. Furthermore, direct calculation gives that 
${}_0{\edth}{}_0{\edth}'\nu^{(2)}={}_0{\edth}'{}_0{\edth}\nu^{(2)}=
-T^{\bi}t_{\bi}$ holds. However, it is precisely the functions $\xi
_{\bi}$ that appear in the BMS rotation vector fields. Indeed, in the 
standard Bondi-type coordinate system $(u,r,\zeta,\bar\zeta)$ the 
general form of the BMS vector fields is 

$$
k^a=\Bigl(H+\bigl(b^{\bi}+\bar b^{\bi}\bigr)t_{\bi}\,u\Bigr)\bigl(
{\partial\over\partial u}\bigr)^a+b^{\bi}{\sqrt{2}\xi_{\bi}\over
1+\zeta\bar\zeta}\bar{\hat m}^a+\bar b^{\bi}{\sqrt{2}\bar\xi_{\bi}
\over1+\zeta\bar\zeta}{\hat m}^a+O\bigl(r^{-1}\bigr), \eqno(4.11)
$$
where $H=H(\zeta,\bar\zeta)$ is an arbitrary real function, and $\hat 
m^a:={1\over\sqrt2}(1+\zeta\bar\zeta)(\partial/\partial\bar\zeta)^a$, 
the Newman--Penrose complex null vector on the unit sphere normalized 
(with respect to the unit sphere metric) such that $\hat m^a\bar{\hat 
m}_a=-1$ (see e.g. [24,25]). Comparing the vector field $N^a$ 
determined by $\nu^{(2)}$ and the BMS vector field above we obtain 
that {\it the vector field $N^a$ corresponding to the function $\nu
^{(2)}$ is the pure rotation BMS vector field with parameters $b
^{\bi}={1\over2}{\rm i}T^{\bi}$}. Thus it seems promising to 
calculate the observable $O[N^a]$ explicitly. It is 

$$
O\bigl[N^a\bigr]={1\over\kappa}\oint_{{\cal S}_1}\Bigl(-k_a\hat m^a
\bigl(\bar\psi^0_{1'}+\bar\sigma^0{}_0{\edth}'\sigma^0\bigr)-k_a\bar
{\hat m}^a\bigl(\psi^0_1+\sigma^0{}_0{\edth}\bar\sigma^0\bigr)+{\rm i}
\sigma^0\bigl({}_0{\edth}'^2\nu^{(1)}\bigr)-{\rm i}\bar\sigma^0\bigl(
{}_0{\edth}^2\nu^{(1)}\bigr)\Bigr){\rm d}{\cal S}_1+O\bigl(r^{-1}
\bigr). \eqno(4.12)
$$
Though the first two terms of the integrand have some resemblance to 
several angular momentum expressions at future null infinity (see e.g. 
[25,26] and references therein), without additional restrictions on 
$\nu^{(1)}$ the last two terms make the whole expression totally 
ambiguous. 

On the other hand, if the spacetime is stationary then the asymptotic 
shear is purely electric: $\sigma^0=-{}_0{\edth}^2S$ for some real 
function $S$ (see e.g. [24]). Bramson [27] showed that in this case 
$2\bar\sigma^0{}_0{\edth}'\sigma^0+{}_0{\edth}'(\sigma^0\bar\sigma^0)
=2{}_0{\edth}'^3\bar A+2{}_0{\edth}\bar B$ for some functions $A$ 
and $B$ built from $S$ and its ${}_0{\edth}$ and ${}_0
{\edth}'$-derivatives. Furthermore, also in the stationary case, 
elementary calculation gives that ${}_0{\edth}'^2\sigma^0={}_0
{\edth}^2\bar\sigma^0$. Substituting these into (4.10) or (4.12), 
and using $k_a\hat m^a={\rm i}{}_0{\edth}(T^{\bi}t_{\bi})$ and 
$T^{\bi}t_{\bi}\in{\rm ker}{}_0{\edth}^2\cap{\rm ker}{}_0{\edth}'
{}^2$, by partial integration we obtain 

$$\eqalign{
O\bigl[N^a\bigr]={1\over\kappa}\oint_{{\cal S}_1}\Bigl(&-k_a\hat m^a
 \bigl(\bar\psi^0_{1'}-{1\over2}{}_0{\edth}'(\sigma^0\bar\sigma^0)+
 {}_0{\edth}'^3\bar A+{}_0{\edth}\bar B\bigr)-\cr
&-k_a\bar{\hat m}{}^a\bigl(\psi^0_1-{1\over2}{}_0{\edth}(\sigma^0\bar
 \sigma^0)+{}_0{\edth}^3A+{}_0{\edth}'B\bigr)\Bigr){\rm d}{\cal S}_1
 +O\bigl(r^{-1}\bigr)=\cr
={1\over\kappa}\oint_{{\cal S}_1}\Bigl(&-k_a\hat m^a\psi^0_1-k_a
 \bar{\hat m}{}^a\bar\psi^0_{1'}\Bigr){\rm d}{\cal S}_1+O\bigl(r^{-1}
 \bigr),\cr} 
\eqno(4.13)
$$
which is the standard spatial angular momentum expression at future 
null infinity [27,25]. Thus in stationary spacetimes the ambiguities, 
coming from the arbitrariness of $\nu^{(1)}$, are cancelled.

\medskip
\ni
{$\bullet$ {\it Large spheres near the spatial infinity}

\ni
Finally suppose that ${\cal S}_r$ is a large sphere of radius $r$ 
near spatial infinity in an asymptotically flat slice. A 
straightforward calculation gives that 

$$
O\bigl[N^a\bigr]=-{1\over\kappa}\oint_{{\cal S}_r}N^a\Pi^b_a\Bigl(
\chi_{bc}-\chi h_{bc}\Bigr)v^c{\rm d}{\cal S}_r=2\int_{\Sigma_r}D_a
\Bigl(\tilde p^{ab}N_b\Bigr){\rm d}^3x=2\int_{\Sigma_r}\Bigl(\bigl(
D_a\tilde p^{ab}\bigr)N_b+\tilde p^{ab}D_{(a}N_{b)}\Bigr){\rm d}^3x, 
\eqno(4.14)
$$
whose $r\rightarrow\infty$ limit is the standard expression of the 
spatial angular momentum for the asymptotic rotation Killing vectors 
$N^a$ [1-3]. However, to have finite and functionally differentiable 
global Hamiltonian the only $N^a$ which is not vanishing at infinity 
must be an asymptotic translation or rotation. Hence by the condition 
$v_aN^a\vert_{{\cal S}_r}=0$ it must be a linear combination of the 
three independent asymptotic rotations. Therefore, {\it at spatial 
infinity $O[N^a]$ reduces to the standard spatial angular momentum}. 

Therefore, to summarize: 
the basic Hamiltonian $H_0[N,N^a]$ is functionally differentiable with 
respect to the canonical variables on each sector $T^*{\cal Q}(\Sigma,
\varepsilon_{a_1...a_{n-1}})$ provided $N$ is vanishing and $N^a$ is 
tangent to ${\cal S}$ on ${\cal S}$. This condition is 
$(n-1)+2$-covariant. If, in addition, $N^a$ is required to be $\delta
_a$-divergence-free on ${\cal S}$, then the boundary conditions on 
the canonical variables are preserved by the evolution equations, the 
basic Hamiltonians form a Poisson algebra in which the constraints 
form an ideal, and the value of the basic Hamiltonian on the 
constraint surface defines a boost gauge-invariant, 
$(n-1)+2$-covariant quasi-local observable associated with the closed 
spacelike $(n-1)$-surface ${\cal S}$. In axi-symmetric spacetimes for 
axi-symmetric surfaces this observable coincides with the Komar 
angular momentum, at spatial infinity it reduces to the spatial 
angular momentum, for small spheres (with the approximate rotation 
Killing fields specified by hand) it is compatible with the expected 
behaviour of a reasonable quasi-local angular momentum expression, 
and in stationary spacetimes it reproduces the standard ambiguity-free 
angular momentum at null infinity. However, without additional 
restrictions on $N^a$ (or on the still freely specifiable function 
$\nu$) it is ambiguous at future null infinity of a radiative spacetime. 
Likewise, for general $\nu$ the integral $O[N^a]$ is not vanishing in 
Minkowski spacetime: that reduces only to the smeared average ${{\rm i}
\over\kappa}\oint_{\cal S}\nu(\bar\sigma\bar\sigma'-\sigma\sigma'){\rm 
d}{\cal S}$ of the two shears of ${\cal S}$. Thus the question arises 
whether we can find conditions on the function $\nu$ for which the 
observable $O[N^a]$ defines ambiguity-free angular momentum at null 
infinity, and, at the quasi-local level, $O[N^a]$ is vanishing in flat 
spacetime. This is still an open question.

\bigskip
\ni
{\lbf Acknowledgments}
\bigskip
\ni
The author is grateful to Robert Beig, Sergio Dain, J\"org Frauendiener, 
Helmut Friedrich, Niall \'O Murchadha, James Nester, Paul Tod, Roh-Suan 
Tung and Jong Yoon for stimulative and helpful discussions, to the 
Morningside Center of Mathematics, Beijing, for hospitality during the 
`Workshop on Quasi-Local Mass', and to the Isaac Newton Institute for 
Mathematical Sciences, Cambridge, where a part of the preset work was 
done. This work was partially supported by the Hungarian Scientific 
Research Fund grant OTKA T042531.

\bigskip
\ni
{\lbf References}
\bigskip
\ni

\item{[1]} T. Regge, C. Teitelboim, Role of surface integrals in the 
           Hamiltonian formulation of general relativity, Ann. Phys. 
           (N.Y.) {\bf 88} 286--318 (1974)
\item{[2]} R. Beig, N. \'O Murchadha, The Poincar\'e group as the 
           symmetry group of canonical general relativity, Ann. Phys. 
           (N.Y.) {\bf 174} 463--498 (1987) 

\item{[3]} L.B. Szabados, On the roots of the Poincar\'e structure 
           of asymptotically flat spacetimes, Class. Quantum Gravity, 
           {\bf 20} 2627--2661 (2003), gr-qc/0302033
\item{[4]} L.B. Szabados, The Poincar\'e structure and the 
           centre-of-mass of asymptotically flat spacetimes, in {\it 
           Mathematical Relativity: New Ideas and Developments}, 
           Eds. J. Frauendiener, D. Giulini and V. Perlick, Springer 
           Lecture Notes in Physics, Springer, Berlin (at press) 

\item{[5]} A.P. Balachandran, A. Momen, L. Chandar, Edge states in 
           gravity and black hole physics, Nucl. Phys. B {\bf 461} 
           581--596 (1996), gr-qc/9412019
\item{[6]} A.P. Balachandran, L. Chandar, A. Momen, Edge states in 
           canonical gravity, gr-qc/9506006v2

\item{[7]} S. Carlip, Statistical mechanics and black hole 
           thermodynamics, gr-qc/9702017

\item{[8]} V. Husain, S. Major, Gravity and BF theory defined in 
           bounded regions, Nucl. Phys. B {\bf 500} 381--401 (1997), 
           gr-qc/9703043

\item{[9]} L.B. Szabados, Quasi-local energy-momentum and angular 
           momentum in GR: A review article, Living Rev. Relativity 
           {\bf 7} (2004) 4, {\tt 
           http://www.livingreviews.org/lrr-2004-4} 

\item{[10]} C.G. Torre, Gravitational observables and local symmetries, 
            Phys. Rev. D {\bf 48} R2373--R2376 (1993), gr-qc/9306030
\item{[11]} C.G. Torre, The problems of time and observables: Some 
            mathematical results, gr-qc/9404029

\item{[12]} C.-M. Chen, J.M. Nester, A symplectic Hamiltonian 
            derivation of quasi-local energy-momentum for GR, Grav. 
            Cosmol. {\bf 6} 257--270 (2000), gr-qc/0001088
\item{[13]} J.M. Nester, General pseudotensors and quasi-local 
            quantities, Class. Quantum Grav. {\bf 21} S261--S280 
            (2004)  

\item{[14]} J.D. Brown, M. Henneaux, On the Poisson brackets of 
            differentiable generators in classical field theory, J. 
            Math. Phys. {\bf 27} 489--491 (1986)

\item{[15]} R.M. Wald, {\it General Relativity}, The University of 
            Chicago Press, Chicago 1984
\item{[16]} F.W. Warner, {\it Foundations of Differentiable Manifolds 
            and Lie Groups}, Graduate Texts in Mathematics No 94, 
            Springer, 1983 

\item{[17]} J.H. Yoon, New Hamiltonian formulation and quasilocal 
            conservation equations of general relativity, Phys. Rev. 
            D {\bf 70} 084037--1-20 (2004), gr-qc/0406047 
\item{[18]} J.D. Brown, J.W. York, Quasilocal energy and conserved 
            charges derived from the gravitational action, Phys. Rev. 
            D {\bf 47} 1407--1419 (1993)
\item{[19]} C.-C.M. Liu, S.-T. Yau, Positivity of quasi-local mass, 
            Phys. Rev. Lett. {\bf 90} 231102--1-4 (2003), 
            gr-qc/0303019 
\item{[20]} A. Ashtekar, B. Krishnan, Dynamical horizons: Energy, 
            angular momentum, fluxes, and balance laws, Phys. Rev. 
            Lett. {\bf 89} 261101--1-4 (2002), gr-qc/0207080

\item{[21]} J. Katz, A note on Komar's anomalous factor, Class. 
            Quantum Grav. {\bf 2} 423--425 (1985)

\item{[22]} L.B. Szabados, On certain quasi-local spin-angular 
            momentum expressions for small spheres, Class. Quantum 
            Grav. {\bf 16} 2889--2904 (1999), gr-qc/9901068

\item{[23]} W.T. Shaw, The asymptopia of quasi-local mass and momentum 
            I. General formalism and stationary spacetimes, Class. 
            Quantum Grav. {\bf 3} 1069--1104 (1986) 
\item{[24]} E.T. Newman, K.P. Tod, Asymptotically flat space-times, 
            in {\it General Relativity and Gravitation: One Hundred 
            Years After the Birth of Albert Einstein}, Vol 2, pp. 
            1--36, Ed. A. Held, Plenum Press, New York 1980 
\item{[25]} L.B. Szabados, On certain quasi-local spin-angular 
            momentum expressions for large spheres near null 
            infinity, Class. Quantum Grav. {\bf 18} 5487--5510 (2001), 
            gr-qc/0109047, Corrigendum: Class. Quantum Grav. {\bf 19}
            2333 (2002) 
\item{[26]} O.M. Moreschi, Intrinsic angular momentum and centre of 
            mass in general relativity, Class. Quantum Grav. {\bf 21} 
            5409--5425 (2004), gr-qc/0209097
\item{[27]} B.D. Bramson, The invariance of spin, Proc. Roy. Soc. 
            Lond. A {\bf 364} 383--392 (1978)

\end